\documentclass[preprint,12pt]{elsarticle}

\usepackage{amssymb}
\def\be{\begin{equation}}
\def\ee{\end{equation}}
\newcommand{\eq}[1]{Eq.\ (\ref{eq:#1})}
\newcommand{\fig}[1]{Fig.\ \ref{fig:#1}}

\journal{Physica A}

\begin{document}

\begin{frontmatter}



\title{Percolation and the pandemic}


\author{Robert M. Ziff}

\ead{rziff@umich.edu}

\address{Center for the Study of Complex Systems and Department of Chemical Engineering, University of Michigan, 
Ann Arbor, Michigan USA 48109-2800}

\begin{abstract}
This paper is dedicated to the memory of Dietrich Stauffer, who was a pioneer in percolation theory and applications of it to problems of society,  such as epidemiology.  An epidemic is a percolation process gone out of control, that is, going beyond the critical transition threshold $p_c$.  Here we discuss how the threshold is related to the basic infectivity of neighbors $R_0$, for trees (Bethe lattice), trees with triangular cliques, and in non-planar lattice percolation with extended-range connectivity.  It is shown how having a smaller range of contacts increases the critical value of $R_0$ above the value $R_{0,c}=1$ appropriate for a tree, an infinite-range system or a large completely connected graph.
\end{abstract}

\begin{graphicalabstract}
 \includegraphics[width=5 in]{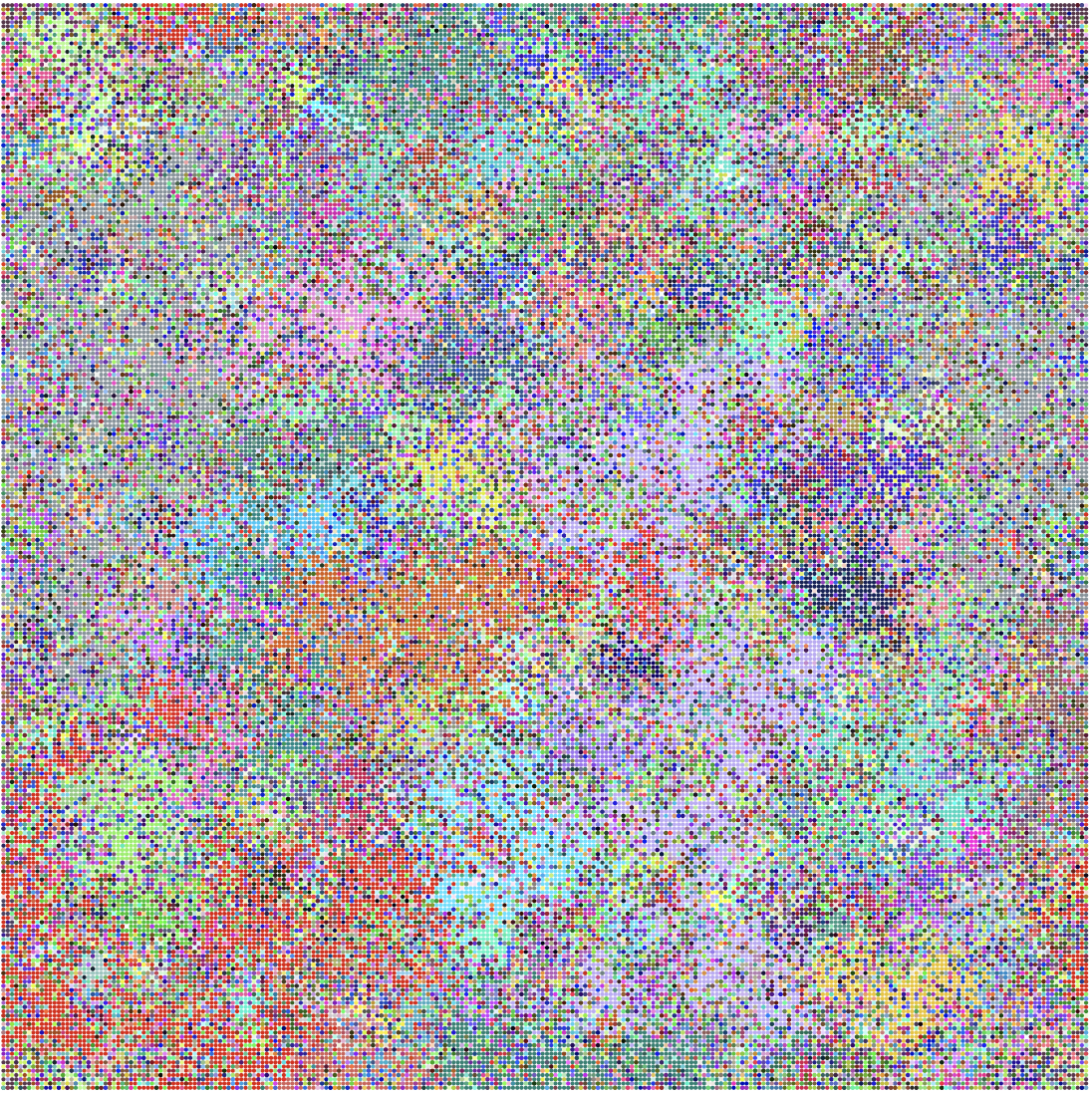}
 
 A system of non-planar extended-range bond percolation with $z=68$ nearest neighbors within a radius of $\sqrt{20}$ around each site, on a square lattice at the bond threshold of $p_c = 0. 019962$.  Different clusters are shown as different colors, and the deep interpenetration of the clusters can be seen.  This system can be considered a model of infection spreading with long but finite-0p-range connections, where different colors represent groups of infected people seeded by different people.
\end{graphicalabstract}

\begin{highlights}
\item Research highlight 1:  For simple tree networks, the threshold corresponds to a critical  infectivity of $R_{0,c} =(\langle z \rangle - 1)p_c = 1$
\item Research highlight 2:  Adding cliques to the tree increases $R_{0,c}$ when considering the outgoing bonds, thus suppressing the formation of an epidemic.
\item Research highlight 3:  For extended-range  site or bond percolation $R_{0,c} > 1$,  which implies clustering with finite-range connections  also suppresses epidemics.
\end{highlights}


\end{frontmatter}


\section{Introduction}

This paper is dedicated to the memory of Dietrich Stauffer, whose work and life had an enormous impact on me.   His encouragement was unflagging, and his criticisms honest.  We had many discussions about aspects of percolation including universality, thresholds, and algorithms over the years, for which I am immensely grateful.

Dietrich was the master of a quickly but clearly written paper,  spontaneous, concise and to the point, something I have always tried to emulate.  When explosive percolation first appeared in with the work of Achlioptas et al.\ \cite{AchlioptasDSouzaSpencer09}, Dietrich suggested to me that I study it using the algorithm of Mark Newman and myself \cite{NewmanZiff00} on the square lattice rather than the complete graph as studied by the authors of \cite{AchlioptasDSouzaSpencer09}.  I told him I was already thinking of it, and he told me it was a here and now problem and that I had to do it immediately---I had one week to get it done!  In fact I finished it in four days and it ended up as a PRL \cite{Ziff09}.  For one week I acted like Dietrich.

One of the projects he repeatedly challenged me to was to find an exact expression for the site percolation threshold on the square lattice, something I had found numerically to relatively high precision ($p_c = 0.592746$)  \cite{ZiffSapoval86,ZiffStell90} and sent to him in time to get it in the second edition of his book with Amnon Aharony \cite{StaufferAharony94}, which allowed this value to be widely known.   I have since worked for several years with Chris Scullard on finding exact thresholds for various lattices in two dimensions, but the square lattice with site percolation cannot be put in the self-dual form that is required to use those methods.  I believe it is insoluble, but you never know---new things come along, such as the isoradial construction \cite{GrimmettManolescu14}, which allowed the inhomogeneous checkerboard critical manifold to be proven \cite{ZiffScullardWiermanSedlock12}.  In recent years, Chris and Jesper Jacobsen have found various thresholds to enormous precision using transfer matrix methods and a percolation criterion related to what had been used on self-dual systems.  For the square-lattice site threshold Jacobsen found $p_c =  0.59274605079210(2)$ \cite{Jacobsen15}, which was independently found to 10 digits in another transfer-matrix calculation \cite{YangZhouLi13}.   This line of work pretty much spells the end of simple Monte Carlo, at least for two dimensions where these methods apply, since, due to statistical error, straight Monte Carlo can never reach that level precision.  (About 8 digits as in \cite{FengDengBlote08} or \cite{MertensMoore18} is the maximum, since this requires at least $10^{16}$ random numbers to be generated.)  When percolation was in its infancy, the best results were found using series, and then Monte Carlo took over.  Now an analytical approach (admittedly with a lot of help from the computer) is again supreme over more brute-force Monte Carlo methods.  

Several years ago there was a meeting of the German Physical Society in Bad Honnef, not far from Cologne where Dietrich lived.  I visited him and he spent two days showing me around the area focusing on the political sites---things related to the war and the post-war period in a divided Germany, which interested him enormously.  It was a wonderful visit.  I didn't know it would be the last time I would see him.

Dietrich was a pioneer in social physics, and in applying ideas related to percolation to various social problems.  Of course a problem related to percolation is epidemiology, a subject of high relevance today.

The basic problem of a spread of a disease is essentially a percolation problem, in which individuals spread an infection to contacts which has the prospect of going out of control and infecting a large fraction of the population---exactly like a percolation transition.

The conventional approach to studying epidemiology is through the  SIR (Susceptible, Infected, Recovered) model of Kermack and McKendrick \cite{KermackMcKendrick27} which dates back almost to the time of the last pandemic.  This model is essentially a mean-field model in which people are put into ``compartments" representing the various categories of Susceptible, Infected, Recovered, and others.  When the basic reproduction number $R_0$, which represents the average number of people ultimately infected by one infected individual, is greater that 1, the infection can grow without bound.  $R_{0,c}=1$  is essentially the percolation threshold for this mean-field model.

When put on a lattice, the SIR model takes into account the geometry of contacts, and is in the same universality class as ordinary percolation \cite{Grassberger83,TomeZiff10}).  Normally one believes that a lattice model cannot be a good model for human disease spread because people travel both locally (work, shopping, entertainment, eating out) and globally (trains, air travel).  During the present pandemic, however, this drawback may not be serious as people are generally  less mobile (although not entirely, of course).  Here, lattice-based immobile models may be more relevant.  In this paper we discuss both Bethe lattice models, which are effectively infinite-dimensional, and regular lattices in two and three dimensions.  
\begin{figure}[htbp] %  figure placement: here, top, bottom, or page
   \centering
   \includegraphics[width=3.4in]{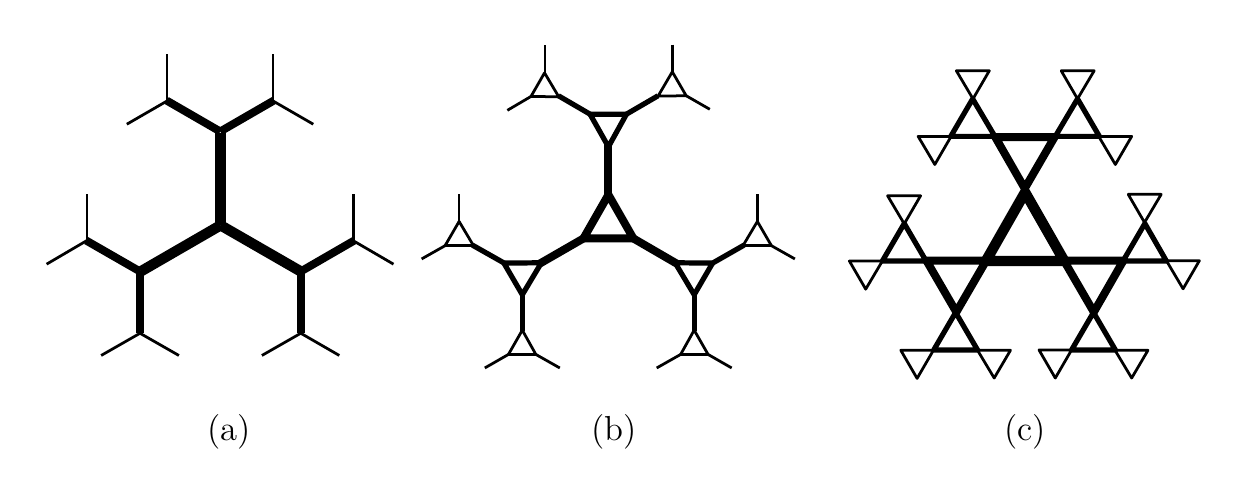} 
   \caption{(a) The Bethe lattice, (b) Husimi tree with bonds, (c) Husimi tree (cactus) of triangles.}
   \label{fig:Pandemicplot3}
\end{figure}

\section{Bethe lattice}

First we review the case where the contacts do not interfere with each other and the infection behavior is essentially as a tree.  This corresponds to the case of a Bethe lattice, or equivalently to low probability on a large fully connected graph (the Erd\H os-R\'enyi  model) where the chance of loop formation is low.  The Bethe lattice is discussed in detail in Stauffer and Aharony's book \cite{StaufferAharony94}.

On a regular Bethe lattice,
each node (site or vertex) is connected to $z$ other nodes in a
tree structure.  An illustration of the Bethe lattice for $z=3$ is shown in \fig{Pandemicplot3}a.
In the bond percolation problem, we assume all bonds (links)
are open or occupied with probability $p$ and closed otherwise.  The bonds represent the transmission 
to the individual who is at the  end of the bond.  The percolation threshold can be found by
the following simple argument: Define $Q$ as the probability that an outgoing
bond from a node does {\it not} connect to infinity (i.e., not be part of the percolating
cluster or giant component).  $Q$ satisfies the equation 
\begin{equation}
Q = 1-p+p Q^{z-1}
\label{eq:BetheQ}
\end{equation}
where the term $1-p$ represents the probability that the bond does not exist, while the second term represents the probability that the bond exists (probability $p$) and the $z-1$ outgoing bonds do not connect to infinity (probability $Q^{z-1}$).  \eq{BetheQ} can be 
written 
\begin{eqnarray}
1-Q &=&p(1-Q^{z-1})\cr &=& p(1-Q)(1+Q+\ldots+Q^{z-2}) 
\label{eq:BetheQ1}
\end{eqnarray}
%Factoring  \eq{BetheQ} one finds
%\begin{equation}
%(Q-1)[1-p(1+Q+\ldots+Q^{z-2}) ]= 0
%\label{eq:BetheQ1}
%\end{equation}
which yields two solutions: $Q = 1$, and the solution for $Q\le1$ to 
\begin{equation}
1+Q+Q^2+\ldots+Q^{z-2} = \frac1p
\label{eq;BetheQ2}
\end{equation}
The transition is  where the two curves intersect, which is at $Q=1$ and $p = p_c$ given by
\be
p_c = \frac{1}{z-1}
\ee
For a Bethe lattice, a spreading epidemic never encounters a previously visited
site, so that all outgoing infections can potentially spread  to the $z-1$ outgoing neighbors, which are all susceptible.  Thus, the average number of the outgoing neighbors that will be infected is 
\be
p_c (z-1) = 1
\ee
which we can associate with the critical value $R_{0,c}$ of $R_{0}=p(z-1)$.  When 
$R_0 > 1$, the epidemic will grow exponentially fast.  Thus, for the simple Bethe
lattice, the critical $R_0$ is equal to 1, consistent with the mean-field result that $R_0>1$ is the criterion for
an epidemic to occur.  

 Of course, this shows that the larger the value of $z$, the probability of infecting a given neighbor $p$ must be decreased to keep the system below the critical point.  But the net number of infected neighbors at the critical point remains 1 no matter what $z$.

The above development can be generalized to systems with a distribution of coordination
number $P_z$, $z = 0, 1, 2, \ldots$ for each node, where $\sum_z P_z = 1$.
The  distribution of neighbors from a node {\it connected to a randomly chosen bond}
is $q_z = z P_z /\langle z \rangle_P$, where $\langle z \rangle_P = \sum_z z P_z$, because a node
with $z$ bonds attached is $z$ times more likely to be chosen than a node with one bond, and we divide by $\langle z \rangle_P$ to normalize the distribution.  The analog to \eq{BetheQ} is now
\begin{eqnarray}
Q &=& 1-p+p \sum_{z\ge1}q_z Q^{z-1} \cr &=& 1-p+p (q_1 + q_2 Q + q_3 Q^2 + q_4 Q^3 \ldots)
\label{eq:BetheQ3}
\end{eqnarray}
By the normalization of $q_z$ we can write
\begin{equation}
q_1 = 1 - q_2 - q_3 \ldots
\label{eq:q1}
\end{equation}
where $q_1$ is the probability that a node reached through a bond has one neighbor, namely
the neighbor where the bond came from, so this is a terminal bond.   Putting \eq{q1} in \eq{BetheQ3} yields
\begin{equation}
1-Q =  p[q_2 (1-Q) + q_3 (1-Q^2) + q_4 (1-Q^3) \ldots]
\label{eq:BetheQ4}
\end{equation}
which has the roots $Q=1$ and the solution to
\begin{equation}
1 = p [q_2 + q_3 (1+Q) + q_4(1+Q+Q^2) +\ldots]
\end{equation}
At $Q=1$, this equation yields the threshold \cite{MolloyReed95}
\be
p_c = \frac{1}{\langle z \rangle_q - 1}
\label{eq:pcBethe}
\ee
where 
\be
\langle z \rangle_q = \sum_{z\ge 1} z q_z = \frac{1}{\langle z \rangle_P} \sum_{z\ge 1} z^2 P_z= \frac{\langle z^2 \rangle_P}{\langle z \rangle_P}
\ee
The quantity $\langle z \rangle_q$ represents the average coordination of nodes connected to a randomly chosen bond.  Again, this threshold corresponds to the point where the outgoing bonds have an average occupancy of one:
\begin{equation}
R_{0,c}  = p_c(\langle z \rangle_q - 1)=1.
\end{equation}

Note that \eq{pcBethe} also can be found by writing $Q = 1 + \epsilon$ and expanding \eq{BetheQ3} about $\epsilon = 0$, since that equation has a double root at $Q = 1$ when you are at the threshold $p_c$.  The relation between $q_z$ and $P_z$ is discussed more in Ref.\ \cite{KryvenZiffBianconi19}.

As a specific example, consider the case where $P_3 = 0.9$ and $P_{10} = 0.1$---that is, 90\% the nodes have coordination number 3 and therefore 2 outgoing neighbors, and 10\%  the nodes have coordination number 10 and therefore 9 outgoing neighbors.  Then $\langle z \rangle_P = (0.9)(3)+(0.1)(10) = 3.7$, $\langle z^2\rangle_P = 18.1$, $\langle z \rangle_q =4.892$ and $p_c =0.2569$ according to \eq{pcBethe}.    If all nodes had coordination number $z=3$, then $p_c$ would be equal to $1/2$.  Thus, having just 10\% of the nodes with $z=10$ reduces the threshold almost in half and greatly increases the risk of an epidemic occurring for a fixed $p$.

Note that in the example above $q_3 = 27/37$ and $q_{10}=10/37$.  Thus, if you pick a site randomly, its average number of neighbors is $\langle z \rangle_P = 3 P_3 + 10 P_{10} = 3.7$,  but if you look at a neighboring site of that  randomly selected site (or the site connected to a randomly selected bond), the average coordination of that site is $\langle z \rangle_q  = 3 q_3 + 10 q_{10}=4.892$---that is, your neighbor is likely to have more neighbors than you have.  This is an example of the phenomenon that your friends are likely to have more friends that you have, and is a general feature of networks where the distribution of neighbors is inhomogeneous.

As another example, we consider the case of a Bethe lattice where the distribution of $z$ around a given site follows a Poisson distribution with mean $\lambda$:
\be
P_z = \frac{\lambda^z}{z!} e^{-\lambda}
\ee
which satisfies $\sum_{z=0}^\infty P_z = 1$.  Then $\langle z \rangle_P = \lambda$, $\langle z^2 \rangle_P = \lambda+\lambda^2$, and $\langle z \rangle_q = \lambda + 1$, and it follows from \eq{pcBethe} that $p_c = 1/\lambda$.  Note that the  average coordination number of a randomly chosen node $\langle z \rangle_P$ and the average coordination number of a node connected to a randomly chosen bond $\langle z \rangle_q$ differ by exactly 1 here; so here it turns out that $p_c = 1/\langle z \rangle_P$, which is not generally the case.
%In this case, $p_c =1/\langle z \rangle_P$, but that is not always true.  [Any particular meaning of this?].

Another way to interpret the above result is that if one creates a network with a Poisson distribution with $\lambda > 1$, and each bond is occupied with probability $p$, then the distribution of occupied bonds would also be a Poisson distribution, but with mean $p \lambda$ and threshold $p_c=1/ \lambda$.   If one makes a network (or Bethe lattice) with a Poisson distribution at $\lambda = 1$, then the system is right at the percolation point when all the bonds are occupied ($p_c = 1$).  If one has a Poisson distribution network with $\lambda < 1$, the system will never percolate.

%Note that for the Poisson distribution, the probability a bond connects to a site of coordination number $z$ is $z P_z/\langle z \rangle_P$, and then the average number of outgoing bonds is $z(z-1) P_z/\langle z \rangle_P$.  The average of this quantity is
%\be
%\sum_{z=0}^\infty z(z-1) P_z /\langle z \rangle_P= \sum_{z=0}^\infty z(z-1)  \frac{\lambda^z}{z!} e^{-\lambda}/\langle z \rangle_P = \lambda
%\ee
%and the critical point, the average number of occupied outgoing bonds is once again $p_c \lambda$.  If we set this equal to $1$ and require that  $p_c = 1$ (all bonds with this distribution are occupied),  then $\lambda=1$, which defines 

The complete graph, or Erd\H os-R\'enyi model, is one in which all $N$ nodes of a system are connected together by $N(N-1)/2$ links, so every node has exactly $N-1$ neighbors.  If we consider occupying a small fraction of those bonds with  $p (N-1)$, the probability of forming loops will be small, and the coordination around each point will be a binomial distribution with mean $p(N-1)$.  In the limit of large $N$ this distribution becomes the Poisson distribution with mean $\lambda = p N$. and  the threshold  will be at $p_c = 1/N$.    
 The ER graph is at the critical point if the mean number of outgoing bonds is equal to 1, just as in the Bethe lattice.
For all examples of a  Bethe lattice, the threshold corresponds to $R_{0,c}  = p_c ({\langle z \rangle_q - 1}) = 1$.

\section{Percolation on the Bethe lattice with internal triangles.}

Here we consider is a variation of the Husimi lattice or a cactus with triangles connected to bonds in 
a tree structure, as shown in \fig{Pandemicplot3}b.  We will consider only a three-branched system here;
the mathematics can be generalized to a larger clique in the center and more
branches.  Such systems have been used in  the study of the spread of diseases  (i.e.,  \cite{IacopiniPetriBarratLatora19}.
First we consider a general system
in which the triangle in the center is a 3-bond (or 2-simplex) characterized by $\mathcal P_0 =$
probability that none of the three vertices are connected together, $\mathcal  P_2=$
the probability that one given pair of vertices are connected together (all 
three possible pairs having equal probability), and $\mathcal  P_3=$
the probability that all three vertices are connected together, such that
\be
\mathcal P_0+3\mathcal P_2+\mathcal P_3=1
\label{eq:norm}
\ee  Then, for $Q$  we have
\be 
Q = 1-p+p(\mathcal P_0+\mathcal P_2+2 \mathcal P_2 Q + \mathcal P_3 Q^2)
\label{eq:husimiQ}
\ee
where $\mathcal P_0+\mathcal P_2$ represents the  events in which the outer
vertices will not connect to the incoming vertex of the triangle,
$2 \mathcal P_2$ represents the two ways there will be one connection,
and $\mathcal P_3$ represents the event in which both outgoing vertices
connect to the incoming one.  % as shown in Fig.\ \ref{fig:husimi}.  

Writing $\mathcal P_0 = 1 - 3 \mathcal P_2 - \mathcal P_3$ and substituting into  \eq{husimiQ}, 
we find
\be
1-Q = p[2 \mathcal P_2 (1-Q) + \mathcal P_3(1-Q^2)]
\ee
which once again has a root at $Q=1$, and a second one at  $1/p = 2 \mathcal P_2 + (1+Q) \mathcal P_3$, and the threshold is determined by the crossing point where $Q = 1$, yielding the criticality condition
\be 
1 = 2p (\mathcal P_2 + \mathcal P_3)
\label{eq:pchusimi}
\ee
where $\mathcal P_2$ and $\mathcal P_3$ can also depend upon internal bond probabilities.

When $p = 1$, we have simply that the threshold is determined by $\mathcal P_2+\mathcal P_3=1/2$, or, using \eq{norm}, the condition
\be
\mathcal P_3 = \mathcal P_0/2 + 1/4
\ee
This can be compared with the condition $\mathcal P_0=\mathcal P_3$ that applies to a regular lattice hypergraph of 3-bonds that fall on a self-dual
arrangement \cite{ZiffScullard06}.
%The point where the two conditions are identical is $P_0 = P_3 = 1/2$, where the system of \fig{Pandemicplot3}c becomes site percolation on the $z=3$ Bethe lattice, while the hypergraph becomes simply site percolation on the triangular lattice with $z=6$.  
The system here is a tree version of the generalized kagome lattice, which has been studied in Refs.\ \cite{ZiffGu09} and \cite{ScullardJacobsenZiff20}, where it was found that  $\mathcal P_3(\mathcal P_0) \approx  0.52732 + 0.83886 (P_0 -  0.096523) + \ldots$ which is an expansion about an exactly soluble point, where the behavior to first-order is also linear.

Now we  consider that the 3-bond is indeed a simple
triangle of three bonds, each occupied with probability $r$.  % while $p$ remains the probability the radial bonds are occupied.
Then $\mathcal P_0 = (1-r)^3$, $\mathcal P_2=r(1-r)^2$ and $\mathcal P_3 = r^3 + 3 r^2 (1-r)$ and \eq{pchusimi} yields the critical curve on the $p$-$r$ plane as
\be 
1 = 2p (r+r^2-r^3)
\label{eq:pchusimi2}
\ee

If $p = r$, that is all bonds on the network have the same probability, then \eq{pchusimi2}  predicts $p_c = 0.637278$.
This is a tree version of the $(3,12^2)$ which on a regular lattice has a threshold of $p_c = [1-2 \sin(\pi/18)]^{1/2}$ =  0.807901 \cite{SudingZiff99}, as expected a  higher value than for the tree.  
For  the two outgoing bonds leaving the triangle, the net $R_{0,c}$ is $2 p_c = 1.615802$, well above the value of 1.  This is a consequence of the clique reducing transmission through it.

When $p = 1$, we have the pure triangular husimi lattice with no bonds linking them together as shown in \fig{Pandemicplot3}c, and the threshold
is determined by \eq{pchusimi2}, with solution $r_c = 0.403032$.  This is a tree version of the kagome lattice, where the bond threshold is 0.524405 \cite{ZiffSuding97,ScullardJacobsen20}, again a higher value.  If we look at a given infected
individual, and consider three out of four of the bonds at a vertex being outgoing bonds, then  $R_0 = 3 r_c = 1.20910$.  This shows that by making the network composed of
closed loops (while overall still a tree), the threshold will correspond to a higher $R_0$, and the closed loops attenuate the probability of an epidemic.

Another model to consider is that either all three of the individuals in the triangular
clique become infected (probability $\mathcal P_3 = r$), or none of them become infected ($\mathcal  P_0 = 1-r$), with $\mathcal P_2 = 0$.  In other words, we are replacing the triangle by a site and making this the site-bond percolation problem on the Bethe lattice.  Here we have from \eq{pchusimi} simply $pr = 1/2$ for the critical curve.  When $r = 1$ we
have the usual $z=3$ Bethe lattice bond threshold $p_c = 1/2$ and $R_{0_c} = 2 p_c = 1$.  Now, however, by lowering
$r$ below 1, we can increase $R_0$ based on the outgoing bonds, $R_{0,c} = 2 p_c = 1/r$.   The minimum value of $r$ is 1/2,
at which point $p_c$ becomes 1 and the system becomes pure site percolation on the Bethe lattice, with $R_0 = 2$.  Having a probability less than 1 of infecting the other people in your group (i.e., decreasing $r$) helps in slowing the epidemic.

While unrelated to the epidemic problem, we note that \eq{pchusimi} is valid for correlated bonds or three-site interactions.  (Of course, the site-bond problem above is also an example of correlated bonds within the triangle.)   For example, consider a system where  $\mathcal  P_0 = 0$, where every triangle has at least one bond in it.  The nomralization condition $3\mathcal  P_2+\mathcal  P_3=1$ and the criticality condition $\mathcal  P_2+\mathcal  P_3 = 1/2$ (\eq{pchusimi} with $p = 1$) implies $\mathcal P_2 = \mathcal P_3 = 1/4$, or it is equally likely that a triangle is in each of the one-bond states or in the fully connected state.   The fully connected triangles triangles ($\mathcal P_3$) act as branches, and the other triangles act as single bonds, or dead-ends if the bond is one opposite side of a triangle from a given path. This is reminiscent to a  model on the regular triangular lattice where each ``up" triangle has exactly one bond, which is also at the critical point \cite{ZiffScullard06}.
% [Not sure of the correct reference here.]

\section{Percolation on lattices with extended-range non-planar connections}

In ordinary bond percolation, the edges or bonds are independently
occupied with probability $p$, and connected nodes (sites) make 
clusters or graph components.  When carried out on a regular lattice,
this leads to the commonly studied models of percolation.

For a spread of an epidemic, a lattice percolation model has the
virtue that it can take spatial effects such as blocking into account.
For most situations, a simple two-dimensional lattice would not
be appropriate, because of longer-range transmission through 
travel etc.  However, in the present Covid-19 pandemic, travel
(both on large and small scales) has been drastically curtailed,
so lattice percolation may possibly be a useful model. 

For a spreading epidemic, we have the same situation as in the Bethe lattice where
the disease enters through one bond.  The $(z-1)$ remaining
bonds are available for spreading.  Assuming there is no other 
infection in this region of the lattice, the average number of bonds that
carry the infection to new sites will be $p (z-1)$.  Thus, if we again
associate $R_0$ with $p(z-1)$,  the critical value of $R_0$
will be
\be
R_{0,c} = p_c (z-1)
\ee
For example, for bond percolation on the square lattice, where $z = 4$
and $p_c = 1/2$, this implies that $R_{0,c}=3/2.$   We see that this is significantly larger than the critical 
value ($R_{0,c} = 1$) of the Bethe lattice, because many of the growth
paths of the epidemic on a lattice are blocked by encounters with previously infected
sites during the growth process.

 However,
it is more relevant  to have a longer range of contact included in 
the  model beyond the nearest neighbor.  This can be incorporated into percolation by adding
long but finite-range bonds to the neighborhood of the individuals, who
sit at the sites of the lattice.  One can think of the longer-range (or extended-range) bonds as representing a limited
amount of mobility for individuals.

The study of extended-range non-planar percolation goes back to the equivalent-neighbor
model of Domb and Dalton \cite{DombDalton66}, where site percolation was studied
with several extended neighborhoods on a variety of lattices.  Other site
percolation models were studied by Gouker and Family \cite{GoukerFamily83}, Galam and Malarz \cite{GalamMauger96},
d'Iribarne et al.\ \cite{dIribarneRasigniRasigni99b}, 
and many others \cite{GawronCieplak91,KurzawskiMalarz12,MajewskiMalarz07,Malarz15,Malarz20,XunZiff20b}.  Here we are more interested in bond percolation, where
an occupied bond represents the spread of an infection to a susceptible 
individual within the allowed neighborhood.   Bond percolation on lattices
with extended neighborhoods has been studied more recently
 \cite{OuyangDengBlote18,DengOuyangBlote19,XunZiff20b,XunHaoZiff20,XuWangHuDeng20}.  Another related 
 model is  long-range percolation which has an infinite range and a power-law bond probability distribution
 \cite{SanderWarrenSokolov2003,Grassberger13,TirnakliTsallis20}.

\setcitestyle{super,open={},close={}}
\begin{table}[htb]
\caption{Bond percolation thresholds for square lattice with circular neighborhoods up to radius $r$, showing the
coordination number $z$ (number of nearest neighbors to a site) and the maximum nearest neighbor (NN) considered.  
Values come from the references cited in superscripts, except $z = 24$, 44, 56, 68 and 80, which we determined here using a single-cluster
growth procedure \cite{XunZiff20b} and a method based upon a universal fluctuation ratio\cite{HuBloteDeng12}.}
\begin{tabular}{c|c|c|c|c}
\hline\hline
 $r^2$  & $z$            & NN & $p_{c}$     & $(z-1)p_c$  \\ \hline
     1      & 4 & 1  &  0.5      &  1.5    \\     
     2      & 8 & 2  &  0.2503685 \cite{OuyangDengBlote18}    &  1.7526    \\  
     4      & 12&  3 & 0.1522203    \cite{OuyangDengBlote18}          &    1.6744  \\ 
     5      & 20& 4  &   0.0841509  \cite{OuyangDengBlote18}      &  1.5989  \\ 
     8      & 24&  5  &   0.067185                                                  &    1.5453    \\
     9      & 28&  6 &  0.0558493 \cite{OuyangDengBlote18}       &  1.5079   \\
   10     & 36& 7  &  0.04169608  \cite{OuyangDengBlote18}   &  1.4594  \\
   13     & 44& 8  & 0.032971  & 1.4178 \\
   16     & 48&  9 & 0.02974268\cite{OuyangDengBlote18} &   1.3979 \\    
   17     & 56&  10  &  0.02489 & 1.3690 \\       
   18     & 60&  11 &   0.0230119\cite{OuyangDengBlote18} & 1.3577 \\    
   20          &  68&  12 &  0.019962  & 1.3373  \\   
   25          & 80 &  13 &   0.016517 & 1.3048 \\   
   49    & 148 & 23 & 0.008342595\cite{OuyangDengBlote18} &  1.2264\\ 
   72   & 224& 32  &0.0053050415(33)\cite{OuyangDengBlote18}  &  1.1830 \\
 225   & 708& 86   &0.001557644(4)\cite{DengOuyangBlote19}  & 1.1013  \\
 389   & 1224& 141  &  0.000880188(90) \cite{OuyangDengBlote18} &  1.0765 \\
 489   & 1652	& 185 &  0.000645458(4) \cite{DengOuyangBlote19} & 1.0657  \\
 961        & 3000& 317 & 0.000349601(3)\cite{DengOuyangBlote19} & 1.0485 \\
 1280        & 4016& 413  &  0.0002594722(11)\cite{OuyangDengBlote18}  & 1.0418  \\
     \hline\hline
\end{tabular}
\label{tab:perc2dbond}
\end{table}
\setcitestyle{numbers}
\setcitestyle{square}

In table \ref{tab:perc2dbond} we show results for $p_c$ (from different sources and calculations here) and $R_{0,c} = (z-1)p_c$ for models
of bond percolation with a neighborhood in the region of a disk of radius $r$, namely
with long range bonds that extend to all sites $(x,y)$ about a given site $(x_0,y_0)$ 
such that $(x-x_0)^2 + (y-y_0)^2 \le r^2$.  Note that for $r > 1$ there are crossing bonds such
that paths of different clusters can cross each other without intersecting, so these models represent non-planar systems.

\begin{figure}[htbp] %  figure placement: here, top, bottom, or page
   \centering
   \includegraphics[width=4in]{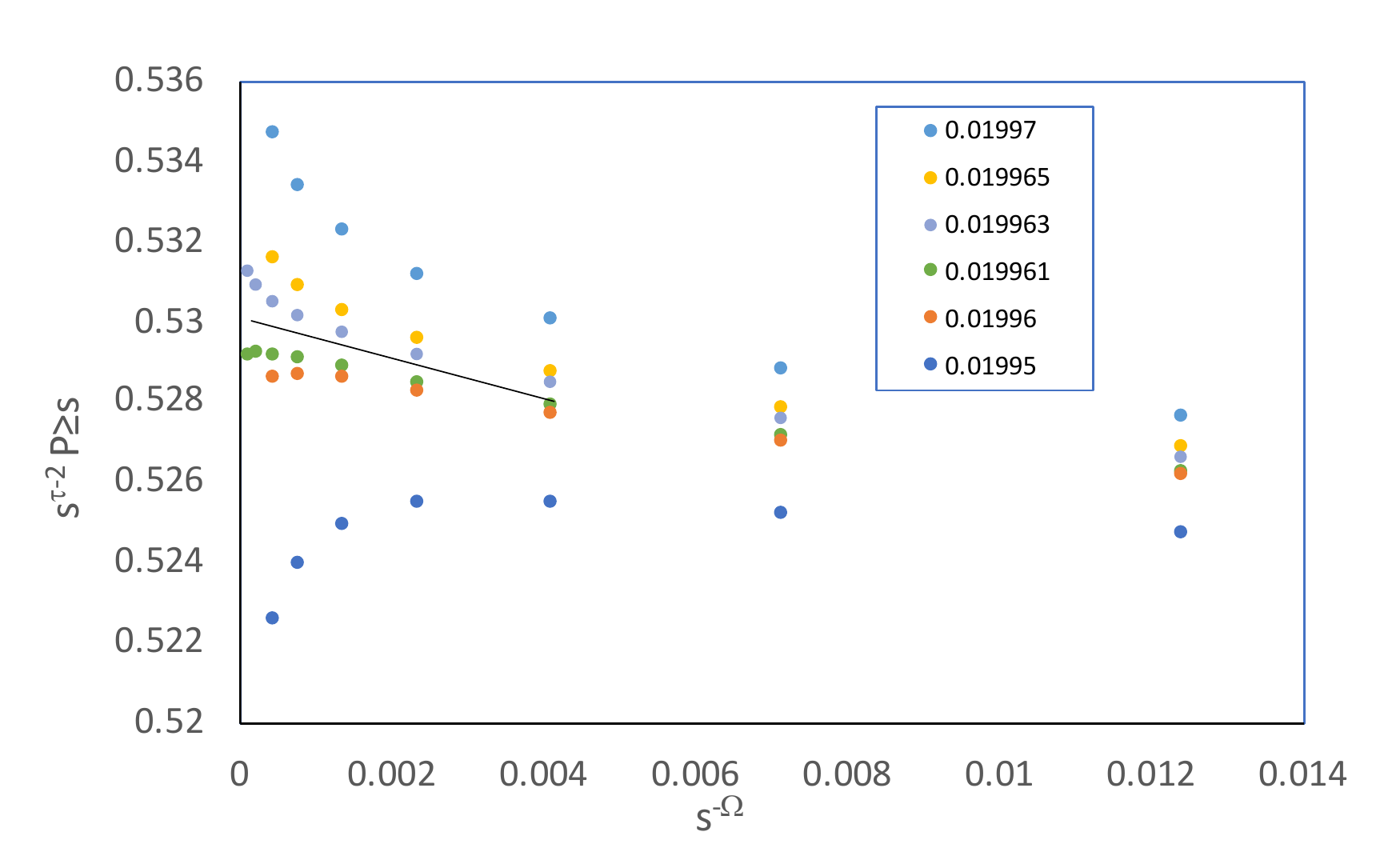} 
   \caption{Determination of the bond percolation threshold for $r = \sqrt{20}$ or $z=68$ using the single-cluster growth algorithm.  A plot of $s^{\tau-2}P_{\ge s}$  vs.\   $s^{-\Omega}$    for different  $p$.  The two central curves suggest a threshold of $p = 0.019962(1)$.
   Here the cutoff was taken as $s_\mathrm{max} = 16384$ except the two central curves with $s_\mathrm{max}=65536$.  Up to $10^{8}$ runs were made for 
   each value of $p$, including the two systems with the larger cutoff. The line goes through the average of the points for $p_c = 0.019961$ and $0.019963$ and shows the expected behavior at the critical point.}
   \label{fig:Pandemicplot4}
\end{figure}

For the calculations here, we used the single-cluster growth method \cite{LorenzZiff98,XunZiff20b} with a maximum cutoff of $s = 16384$ or $65536$ occupied sites, and used the fact that near $p_c$, the probability a site is connected to a cluster of size greater than or equal to $s$ is expected to behave as 
\begin{equation}
P_{\ge s} \sim A s^{\tau-2}[1 + B(p-p_c)s^\sigma+C s^{-\Omega}]
\end{equation}
where $\tau-2=5/91$, $\sigma = 36/91$, and $\Omega=72/91$ in 2D.  This is valid for $p$ very close to $p_c$ and $s$ large.  At $p_c$, a plot of $s^{\tau-2} P_{\ge s}$ vs.\ $s^{-\Omega}$ yields a straight line, as shown in Fig.\ \ref{fig:Pandemicplot4} for $z=68$.  Note that we are finding consistency with the usual 2D value of $\tau$ here; for very large $z$, one might expect Bethe-lattice behavior $\tau = 5/2$ for smaller $s$, but we did not see that here.  

We also verified this result by using the union-find simulation method \cite{NewmanZiff01}, monitoring the largest cluster $s_{max}$ and its square, and finding their average values and the  ratio $Q = \langle s_\mathrm{max}^2 \rangle/\langle s_\mathrm{max} \rangle^2$.  This quantity has the universal value of 1.04149 in 2D, for systems with a square periodic boundary \cite{HuBloteDeng12}.  In a  shorter amount of time than in the single-cluster simulations and in a single run, taking a lattice of size $1024\times1024$ and using $10^6$ samples, we were able to verify the value $p_c = 0.0199616(2)$ as the point where $Q = 1.04149$ for $L = 2048$ (270,000 samples).   This method was also used for finding $p_c$ for $z =$ 24, 44, 56, and 80 also.  These simulation results support that non-planar percolation is in the normal universality class of ordinary percolation, including the ordinary critical exponents and universal fluctuation ratio $Q$.

\begin{figure}[htbp] %  figure placement: here, top, bottom, or page
   \centering
   \includegraphics[width=3.4in]{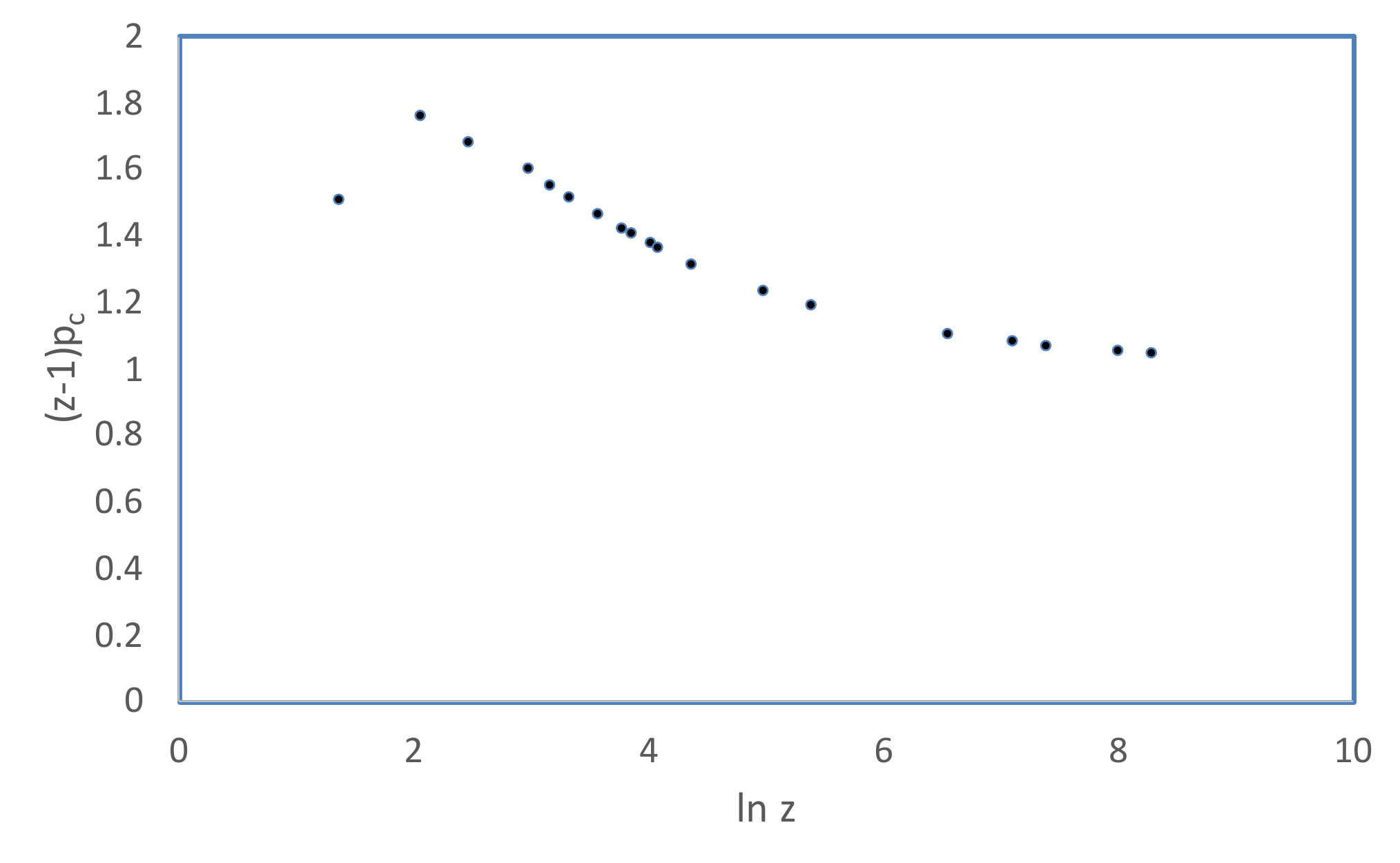} 
   \caption{Plot of $R_{0,c}=(z-1)p_c $ vs.\ $\ln z$ using the data of table \ref{tab:perc2dbond} (2D percolation) }
   \label{fig:Pandemicplot1}
\end{figure}

In \fig{Pandemicplot1} we plot $R_{0,c}=(z-1)p_c$ as a function of $\ln z$,  For large $z$, the asymptotic
behavior $R_{0,c}=1$ is obtained, but for smaller $z$ there is a marked increase, with a peak of $R_{0,c} = 1.7526$ at $z = 8$ but then drops off fairly slowly with $z$.  This shows that 
by restricting possible contacts to a finite range, the critical infection rate will be substantially larger than 1, so that it is less likely 
for an epidemic to spread for a given $R_0$.

\begin{figure}[htbp] %  figure placement: here, top, bottom, or page
   \centering
   \includegraphics[width=3.4in]{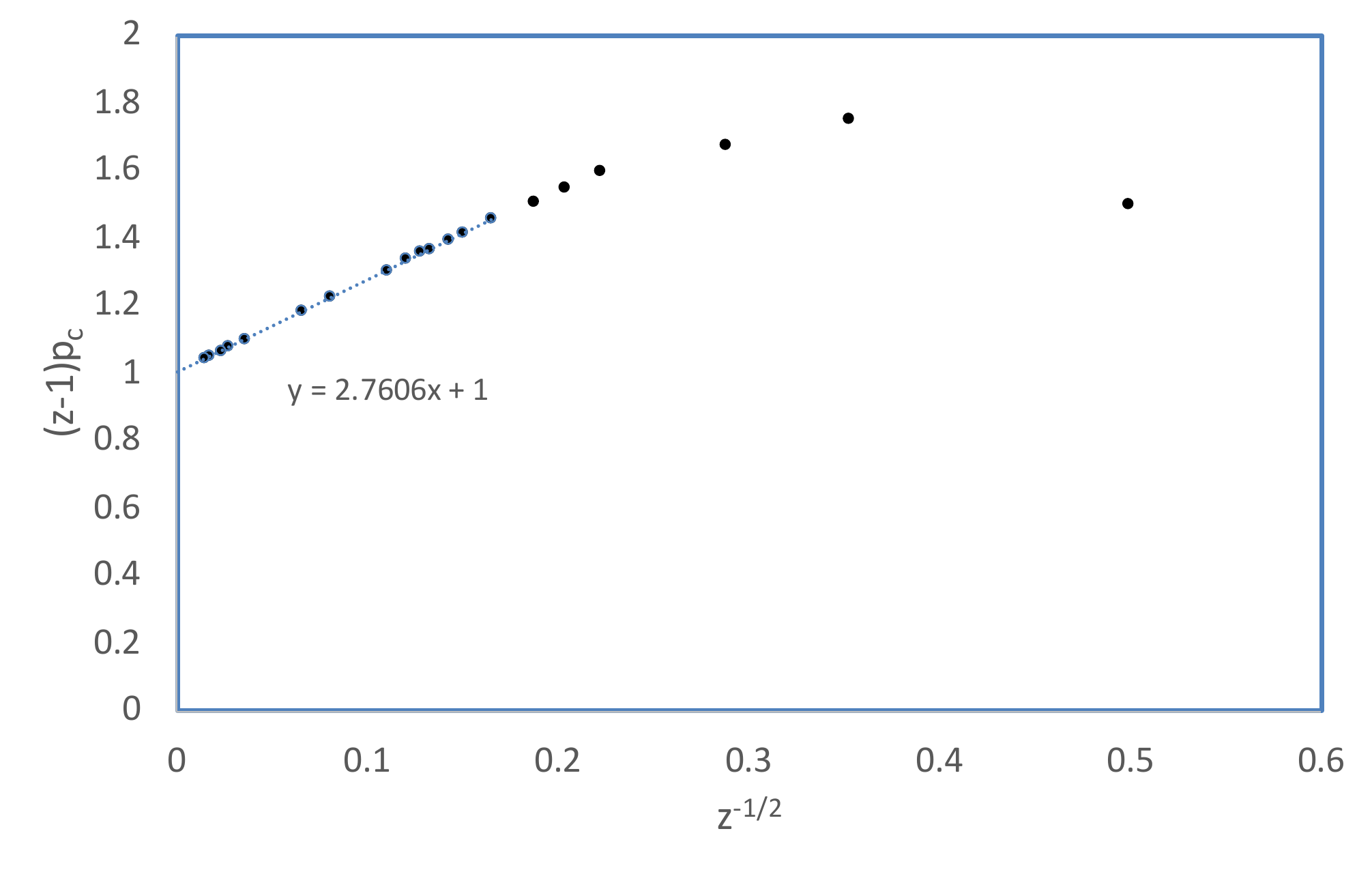} 
   \caption{Plot of $R_{0,c}=(z-1)p_c $ vs.\ $z^{-1/2}$ using the data of table \ref{tab:perc2dbond} (2D percolation).  The linear fit for the
   data for $z \ge 36$ yields a slope  of 2.76.}
   \label{fig:Pandemicplot2}  
\end{figure}

Note that the large-$z$ behavior of $p_c$ for bond percolation is expected to behave as \cite{FreiPerkins16}
\be
(z-1) p_c(z) \sim 1 + c z^{-1/2}
\label{eq:z1pc}
\ee
(The authors of Ref.\ \cite{FreiPerkins16} consider $z p_c$ rather than $(z-1)p_c$, but to this order in $z$ the behavior is the same.)  In \fig{Pandemicplot2} we plot $(z-1)p_c$ as a function of $z^{-1/2}$, and it clearly can be seen that the large-$z$ behavior satisfies this expected form, with $c = 2.756$.  (A related plot is given in Ref.\ \cite{XuWangHuDeng20}, Fig.\ 9, which also verifies the same asymptotic behavior.)   Note that the neighborhoods within a distance of radius $r$ are not exactly the same shape, because of the lattice effect, so one might  expect some deviations from a smooth behavior.  Nevertheless, the adherence to linear behavior on this plot is quite good.
In Ref.\ \cite{FreiPerkins16}, in Corollary 1.3, the authors reference results of a computer simulation that implies $c = 1.2\sqrt{\pi} = 2.13$, which is a bit below the value here and the difference is likely a result of a lower level of statistics in that work.

\begin{figure}[htbp] %  figure placement: here, top, bottom, or page
   \centering
   \includegraphics[width=3.4in]{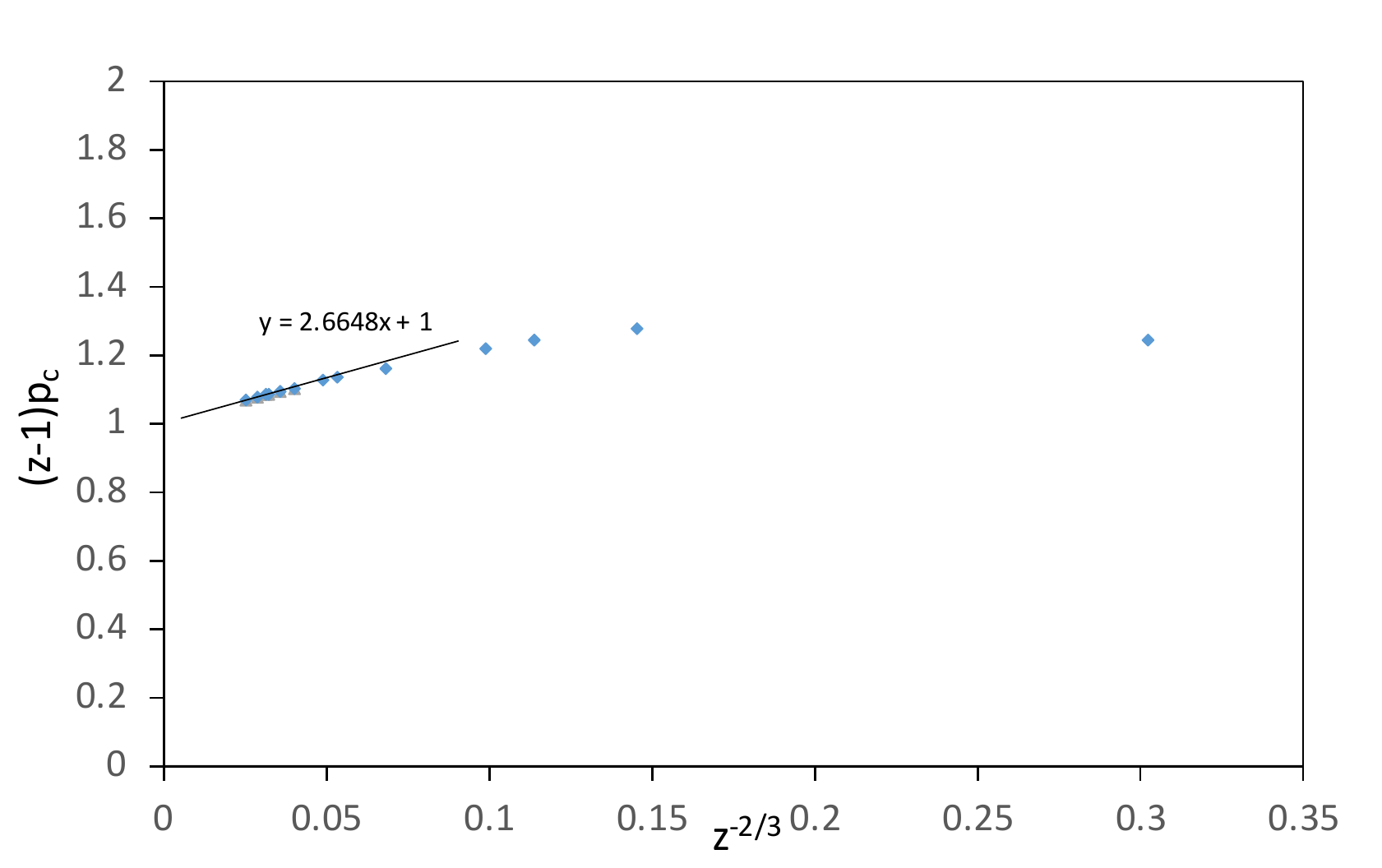} 
   \caption{Plot of $R_{0,c}=(z-1)p_c $ vs.\ $z^{-2/3}$ for the 3D extended-range bond percolation data of Refs.\ \cite{XunZiff20b} and \cite{XunZiff21}.  The peak in $R_0 = (z-1)p_c$ is lower here than the peak in 2D systems.  The line with slope 2.665 fits the six points of data from $z=146$ to 250, and supports the prediction that the corrections go as $z^{-1/3}$}
   \label{fig:Pandemicplot5}  
\end{figure}

In Fig.\ \ref{fig:Pandemicplot5} we make a similar plot for 3D extended-range bond percolation, using data from Ref.\ \cite{XunZiff20b} for smaller $z$ and from more recent work \cite{XunZiff21} for larger $z$ up to $z = 250$.  We find the data are consistent with a correction of order $z^{-2/3}$ as predicted in Ref.\ \cite{FreiPerkins16}, providing a better fit than assuming a correction of $z^{-1/2}$ as in 2D (Eq.\ (\ref{eq:z1pc})) and was used in our previous paper \cite{XunZiff20b}.   Further work for even larger $z$ is needed to demonstrate the correction behavior clearly.

\section{Epidemics and extended-range site percolation}
 
In extended-range site percolation with circular neighborhoods, all occupied sites that are within a circle of radius $r$ of a given occupied site will be connected to that site through long bonds of length greater than one lattice spacing.  In terms of epidemiology, site percolation corresponds to saying that an infected individual will infect all susceptible neighbors within a radius $r$, and the rest of the sites within that radius will never become susceptible or infected.   

Extended-range site percolation differs from extended-range bond percolation in that the former is effectively planar while the latter can be non-planar --- that is, paths of clusters can cross each other in extended-range bond percolation, but cannot in extended-range site percolation.  In site percolation, one assumes that infected individual will  infect all the people within the range of interaction and there are no additional susceptible individuals within that range, which is not the case for bond percolation.

For large $r$, the extended-range site percolation becomes equivalent to the continuum percolation of overlapping disks of radius $r/2$, because two disks of that radius will just touch when their centers are a distance $r$ apart, and the centers of the disks is where we consider the infected individuals to reside.  In continuum percolation of disks, the critical coverage fraction of disks of radius $r/2$ (including overlapping areas) is defined as
\be
\eta_c = \pi (r/2)^2 \frac{N}{V} 
\ee
where $N$ is the number of disks in a region of area $V$ and $\eta_c = 1.12809$ \cite{XuWangHuDeng20,TarasevichEserkepov20,MertensMoore12}.  Then  $p = N/V$ is effectively the occupation of sites when the continuum percolation is superimposed on a lattice.     The number of possible neighbors on the lattice is $z = \pi r^2$, and putting this all together yields \cite{XunZiff20b}
\be
z p_c = 4 \eta_c = 4.5123
\ee
so once again the threshold is inversely proportional to $z$, but with a different coefficient than in bond percolation, where the coefficient is 1 as in  \eq{z1pc}.  From the epidemic perspective, this means that you can infect up to 4.5 people in your circle of influence and not have the epidemic spread.  This is because those and subsequent people will have a smaller area of susceptible people to infect, because of the overlap of the disks of influence.  While in bond percolation, each individual has to infect just a little over one other individual to trigger an epidemic, but eventually more people within the circle of radius $r$ might eventually become infected by other bonds.  It is an interesting question for future study to ask how many individuals within that radius $r$ of the first site eventually become infected through the bond percolation process, by means for direct paths and indirect paths.

Thus, the site percolation result says that if people keep fewer than $R_{0,c} = 4.5123$  other people within their circle, the epidemic will not spread (assuming that the people are spread randomly over a two-dimensional surface). 
%\section{SIR on a lattice}

\section{Discussion}

For a mean field or Bethe-tree behavior, one would expect that the critical value of the infection number $R_0$ is 1.  Putting
groups of people in a clique or otherwise making the tree have loops increases the critical $R_0$ above 1.   Furthermore, putting
the individuals on a plane such as a square lattice with a limited range of infection also has the effect of increasing the critical $R_{0,c}$ above the
value of 1, considering both site and bond percolation models. Clearly, from a public health perspective, clustering people and restricting the range that people can interact can inhibit the spread of a disease, and we can see that percolation theory provides a quantitative explanation of this.  

It should be emphasized that in spite of the observation that for many systems $R_{0,c}>1$, the mean-field SIR model with $R_{0,c} = 1$ remains very useful as a simple set of differential equations for understanding how various factors affect the spread of a disease and for easily finding the time dependence.

Here we assume the individuals are stationary but interact over possibly long but finite range bonds.  The universality class of these models is still that of ordinary percolation.  It is possible that the class may change if movement of the individual is also added to the models, as was studied in \cite{deOliveira21}.

Here we considered just a square lattice.  It may also be interesting to look at a triangular lattice or even a random one
such as the Voronoi lattice or the Delaunay triangulation, although one would expect the behavior to be qualitatively similar.
Higher dimensional lattices might be useful to model more complex interactions.  A direct study of lattice versions of the
SIR model with extended-range interactions should also be interesting.  Another question to look at are the dynamics on various
lattice structures such as small-world and other fractal networks.  This might explain the power-law growth behavior seen in some regimes
of Covid epidemics \cite{TirnakliTsallis20,ZiffZiff20,BianconiKrapivsky20}.

\section{Acknowledgments}

The author thanks Ivan Kryven for comments, and Chris Scullard for comments and for preparing Fig.\ \ref{fig:Pandemicplot3}.

\bibliographystyle{apsrev4-1}
\bibliography{bibliography20.bib}
\end{document}